\documentclass[twocolumn]{revtex4}

\usepackage{graphicx}

\newcommand{\up}{$\uparrow$\ }
\newcommand{\down}{$\downarrow$\ }
\newcommand{\upg}{$| \uparrow g \rangle$}
\newcommand{\downg}{$| \downarrow g \rangle$}
\newcommand{\upe}{$| \uparrow e \rangle$}
\newcommand{\downe}{$| \downarrow e \rangle$}
\newcommand{\uph}{$| \uparrow h \rangle$}
\newcommand{\downh}{$| \downarrow h \rangle$}
\newcommand{\e}{$| e \rangle$}

\newcommand{\units}[1]{\ensuremath{\mathrm{#1}}}
\newcommand{\amount}[2]{\ensuremath{#1\:\units{#2}}}
\newcommand{\sym}[2]{\ensuremath{#1_{\mathrm{#2}}}}

\begin{document}

\title{Tunable spin-selective loading of a silicon spin qubit}

\author{C. B. Simmons, J. R. Prance, B. J. Van Bael, Teck Seng Koh, Zhan Shi,
  D. E. Savage, M. G. Lagally, R. Joynt, Mark Friesen, S. N. Coppersmith,
  and M. A. Eriksson}
\affiliation{University of Wisconsin-Madison, Madison, WI 53706}

\maketitle

\textbf{
The remarkable properties of silicon have made it the central material for the fabrication of current microelectronic devices. Silicon's fundamental properties also make it an attractive option for the development of devices for spintronics \cite{Appelbaum:2007p2644}  and quantum information processing \cite{Kane:1998p133,Loss:1998p120,Vrijen:2000p1643,Friesen:2003p121301}.  The ability to manipulate and measure spins of single electrons is crucial for these applications. Here we report the manipulation and measurement of a single spin in a quantum dot fabricated in a silicon/silicon-germanium heterostructure. We demonstrate that the rate of loading of electrons into the device can be tuned over an order of magnitude using a gate voltage, that the spin state of the loaded electron depends systematically on the loading voltage level, and that this tunability arises because electron spins can be loaded through excited orbital states of the quantum dot. The longitudinal spin relaxation time $T_1$ is measured using single-shot pulsed techniques~\cite{Elzerman:2004p431} and found to be $\sim 3$ seconds at a field of 1.85~Tesla. The demonstration of single spin measurement as well as a long spin relaxation time and tunability of the loading are all favorable properties for spintronics and quantum information processing applications.
}

Silicon is a material in which spin qubits are expected to have long coherence
times, thanks to the predominance of a spin-zero nuclear isotope and
relatively weak spin-orbit coupling.
However, silicon quantum dots have yet to
demonstrate the reproducibility and controllability achieved in gallium arsenide devices \cite{Petta:2005p2180,Koppens:2006p766,PioroLadriere:2008p776,Amasha:2008p1500}.
Here, we demonstrate the control and manipulation of spin states of single
electrons in a silicon/silicon-germanium (Si/SiGe) quantum dot and
report the first single-shot measurements of the longitudinal spin
relaxation time ${ T_1}$ in such devices.
We also show that the
presence of a relatively low-lying spin-split orbital excited state in the dot
can be exploited to increase the speed and tunability of the
loading of spins into the dot.
Our results demonstrate that Si/SiGe quantum dots
can be fabricated that are sufficiently tunable to enable
single-electron manipulation and measurement, and that
long spin relaxation times are consistent with the orbital and/or valley excitation
energies in these systems.

\begin{figure}
\includegraphics[width=8.5cm]{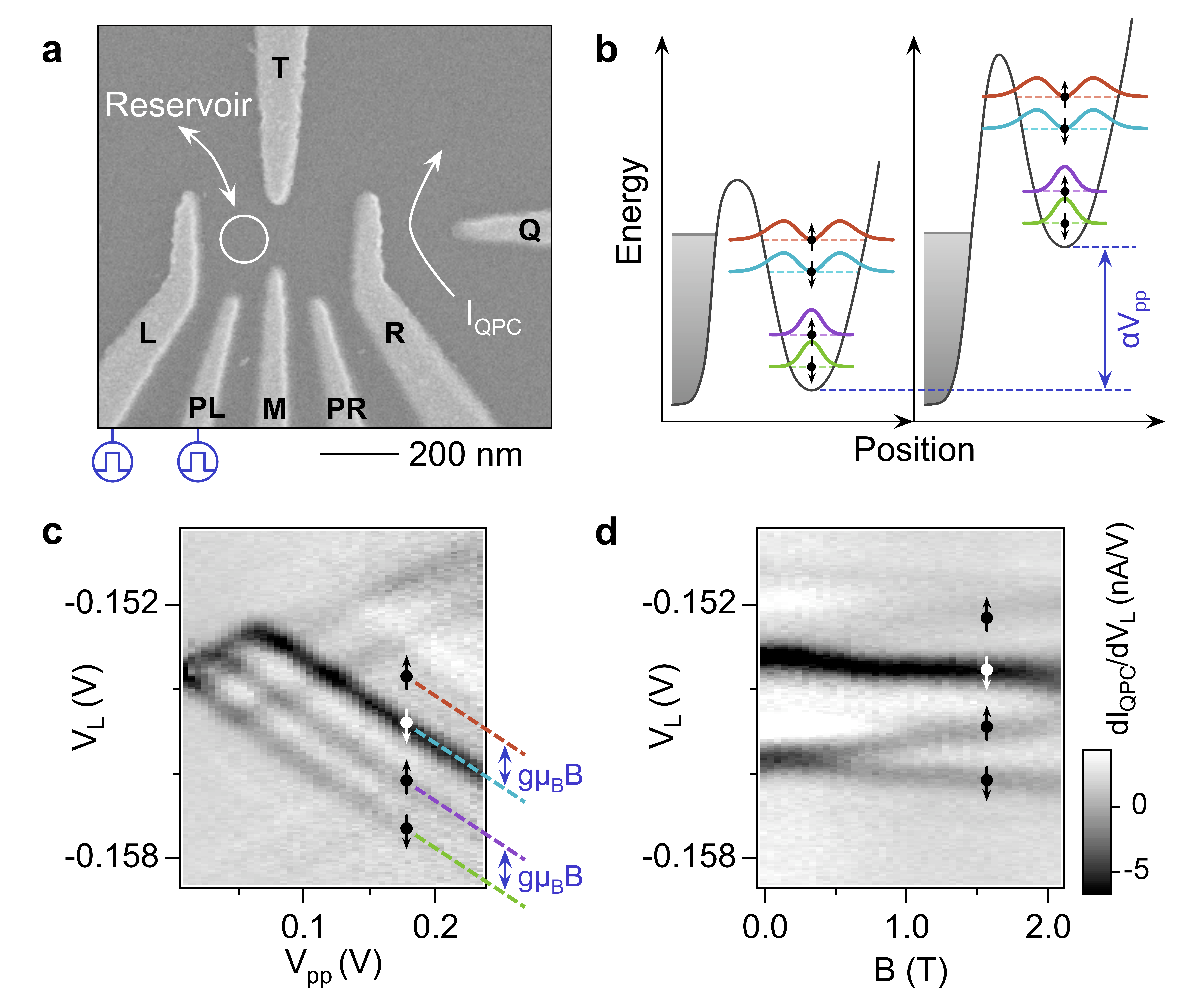}
\caption{\label{fig:expt}
\textbf{Device and spin-state spectroscopy.} \textbf{a,} Scanning electron micrograph of a device identical to that used here.   The single quantum dot location is indicated by the white circle. The dot barriers are tuned so that electrons can tunnel to and from the left reservoir only, and charge sensing is performed by measuring the current labeled \sym{I}{QPC}.
 \textbf{b,} Voltage pulses of amplitude \sym{V}{pp}, applied to either gate L or PL, adjust the energy levels of the dot.  The case shown corresponds to the possibility of loading an electron during the positive phase of the pulse (left) through any of the four states \upg, \downg, \upe, \downe~that are below the Fermi energy $E_F$, and then unloading the electron during the negative phase of the pulse (right), a process that will always occur from \upg~or \downg, because of the fast orbital decay rate. 
\textbf{c \& d,} Measurements of $d\sym{I}{QPC}/d\sym{V}{L}$, in the presence of a pulsed voltage \sym{V}{pp} on gate L, as a function of \sym{V}{L} and the amplitude \sym{V}{pp} (shown in \textbf{c}) or the magnetic field $B$ (shown in \textbf{d}).  The three lowest eigenstates, \downg, \upg, and \downe, are clearly visible in \textbf{c}, where the magnetic field \amount{B=1.5}{T}.  The state \upe~is expected at the location shown by the red dashed line but is invisible due to the strong coupling of the excited state \e~(see Supplemental information).  As a function of $B$ (and with \amount{\sym{V}{pp}=0.125}{V}), the ground and first orbital excited states split linearly due to the Zeeman effect.
}
\end{figure}

The measurements we report were performed on a gate-defined quantum dot with the gate configuration shown in Fig.~\ref{fig:expt}a, tuned to be in the single-dot regime.  The dot is measured at low temperature and in a parallel magnetic field.  As shown in Fig.~\ref{fig:expt}b, an electron can be loaded into one of four energy eigenstates; we denote the states, in order of increasing energy, as \downg, \upg, \downe, and \upe, where the first index refers to spin (\down having lower energy than \mbox{\up)} and the second to the ground ($g$) and excited ($e$) orbital levels.  We obtain an experimental map of these states by measuring the differential current $d\sym{I}{QPC}/d\sym{V}{L}$ through a charge sensing quantum point contact while applying square voltage pulses to gate L.  The grayscale plots of $d\sym{I}{QPC}/d\sym{V}{L}$ in Fig.~\ref{fig:expt}c,d reveal dark lines when energy levels on the dot come into resonance with the Fermi energy \sym{E}{F}~\cite{Elzerman:2004p731}.  The measured calibration factor for gate L, \amount{\alpha = 0.125 \pm 0.006}{eV/V} (see Supplemental information), translates the positions of these lines into a spectroscopy of the dot energy levels.

In Fig.~\ref{fig:expt}c, the darkest line is the orbital excited state \e~of energy \amount{311 \pm 19}{\mu eV}.  This relatively large energy splitting is favorable for applications in which spin coherence is desirable~\cite{Tahan:2002p035314}, and it is notable because
the low-lying excited orbital states in Si/SiGe quantum dots have fundamental differences compared to those in GaAs~\cite{Hanson:2007p1217}, because of the role of valley degrees of freedom~\cite{Boykin:2004p115,Goswami:2007p41,Kharche:2007p092109,Lansbergen:2008p1545,Culcer:2009p205302,Friesen:2010p115324,Fuechsle:2010p502}.  The Zeeman splitting of the ground and excited states is shown in Fig.~\ref{fig:expt}d.

\begin{figure}
\includegraphics[width=8.5cm]{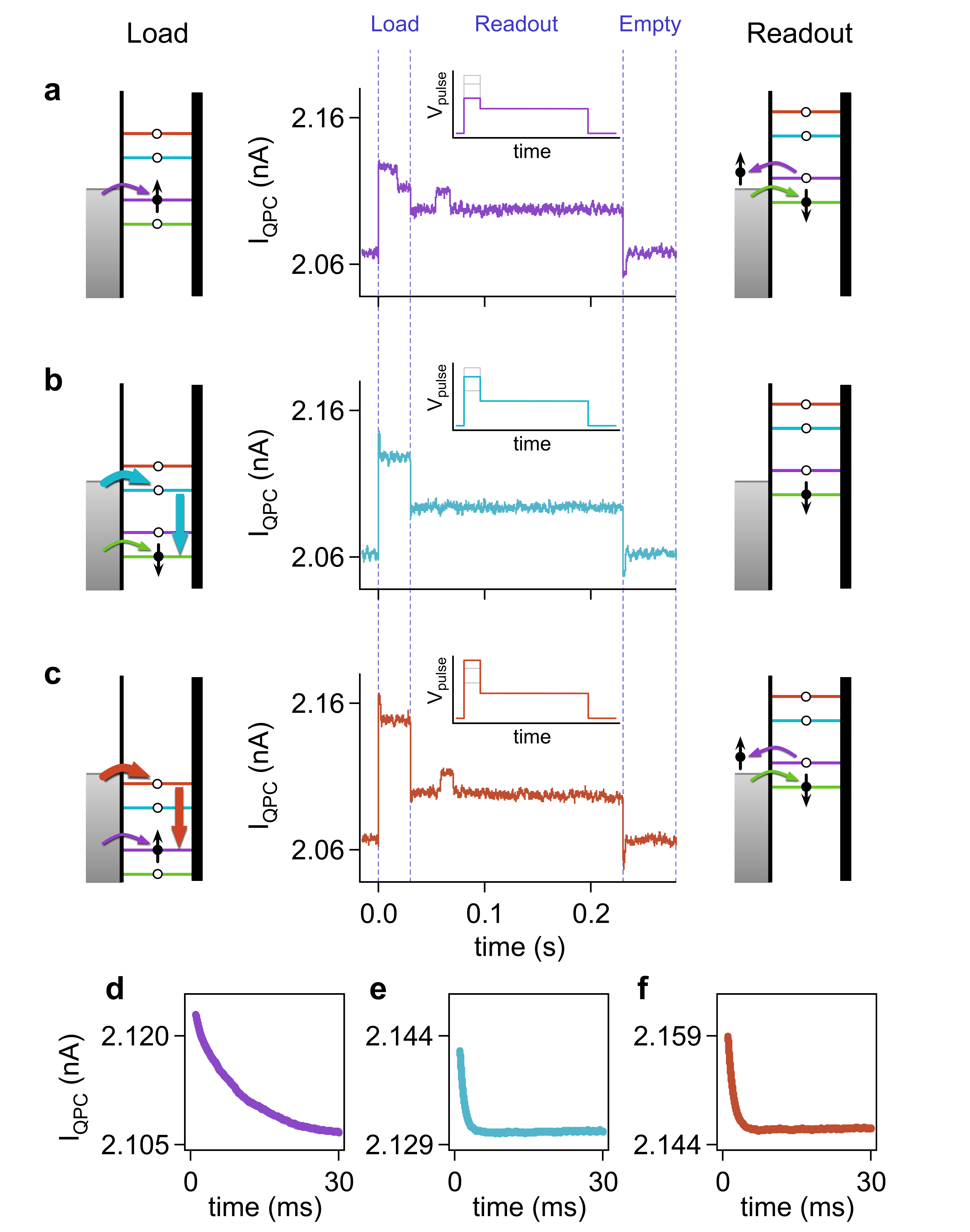}
\caption{\label{fig:measurement}\textbf{Single-shot spin readout and tunable loading rates.} 
Measurement of the spin state of an electron loaded into the dot is performed by applying
a 3-level pulse sequence to gate PL~\protect{\cite{Elzerman:2004p431}}.
Starting with the electron unloaded, \sym{V}{PL} is rapidly changed to a load level.  Panels \textbf{a}, \textbf{b}, \textbf{c} correspond respectively to preferential loading of states \upg, \downe, and \upe~at load voltages $\sym{V}{load}=+175$, $+325$, and \amount{+425}{mV}.  \sym{V}{PL} is then changed to the readout level (\amount{+75}{mV}), where the energy of a spin-\up electron is above \sym{E}{F}, whereas the energy of a spin-\down electron is below \sym{E}{F}.  During the readout stage, a spin-\up electron will tunnel off the dot, and a spin-\down electron will tunnel on to replace it, resulting in the current pulse shown in the readout phase for \textbf{a} and \textbf{c}.  In contrast, a spin-\down electron will remain in the dot and no current pulse will occur during the readout phase, as shown in \textbf{b}.
During the empty stage, the electron tunnels off the dot regardless of its spin orientation.
For all cases: \amount{\sym{t}{load}=30}{ms}, \amount{\sym{t}{readout}=200}{ms}, \amount{\sym{t}{empty}=50}{ms}, and \amount{\sym{V}{empty}=-200}{mV}.  
\textbf{d},  \textbf{e}, \textbf{f} Average of 500 time traces of electron loading events for the load voltages of panels \textbf{a}, \textbf{b}, and \textbf{c}, respectively.  Loading at the excited orbital state (shown in both \textbf{e} and \textbf{f}, with tunnel rates of \amount{1186 \pm 3}{Hz} and \amount{958 \pm 2}{Hz}, respectively), is much faster than loading at the ground state (shown in \textbf{d}, with a tunnel rate of \amount{106.7 \pm 0.2}{Hz}).  The tunnel rates and uncertainties come from exponential fits.
}
\end{figure}

A main focus of this work is to demonstrate that orbital excited states can be
exploited to control both the rate of loading and the spin state of the electron loaded into the dot. To measure the spin state of individual electrons, we use a pulsed-gate technique, pioneered in Ref.~\cite{Elzerman:2004p431} and summarized in Fig.~\ref{fig:measurement}a. An electron is loaded into either the spin-\down or spin-\up ground states, \downg~or \upg.  Spin readout is performed by shifting only \upg~above the Fermi level, so that a spin-\up electron will tunnel out of the dot, but a spin-\down electron will remain.  When a spin-\up exits the dot, a spin-\down will reload, because \downg~is below \sym{E}{F}.  The readout signal is a pulse in \sym{I}{QPC} that persists as long as the dot is unloaded.

We show in Fig.~\ref{fig:measurement}b, c that loading also can be performed into the spin-split excited states, \downe~and \upe, and that loading into these states is much faster than loading into the ground state. The loading rate can be determined by averaging many loading events together, resulting in an exponential decay of \sym{I}{QPC} from a magnitude corresponding to an unloaded electron to that corresponding to a loaded electron.  Panels d--f show that the loading rates into the excited orbital states \mbox{\downe}~and \upe~are much faster than that into the spin-\up ground state \upg.

\begin{figure}
\includegraphics[width=8.5cm]{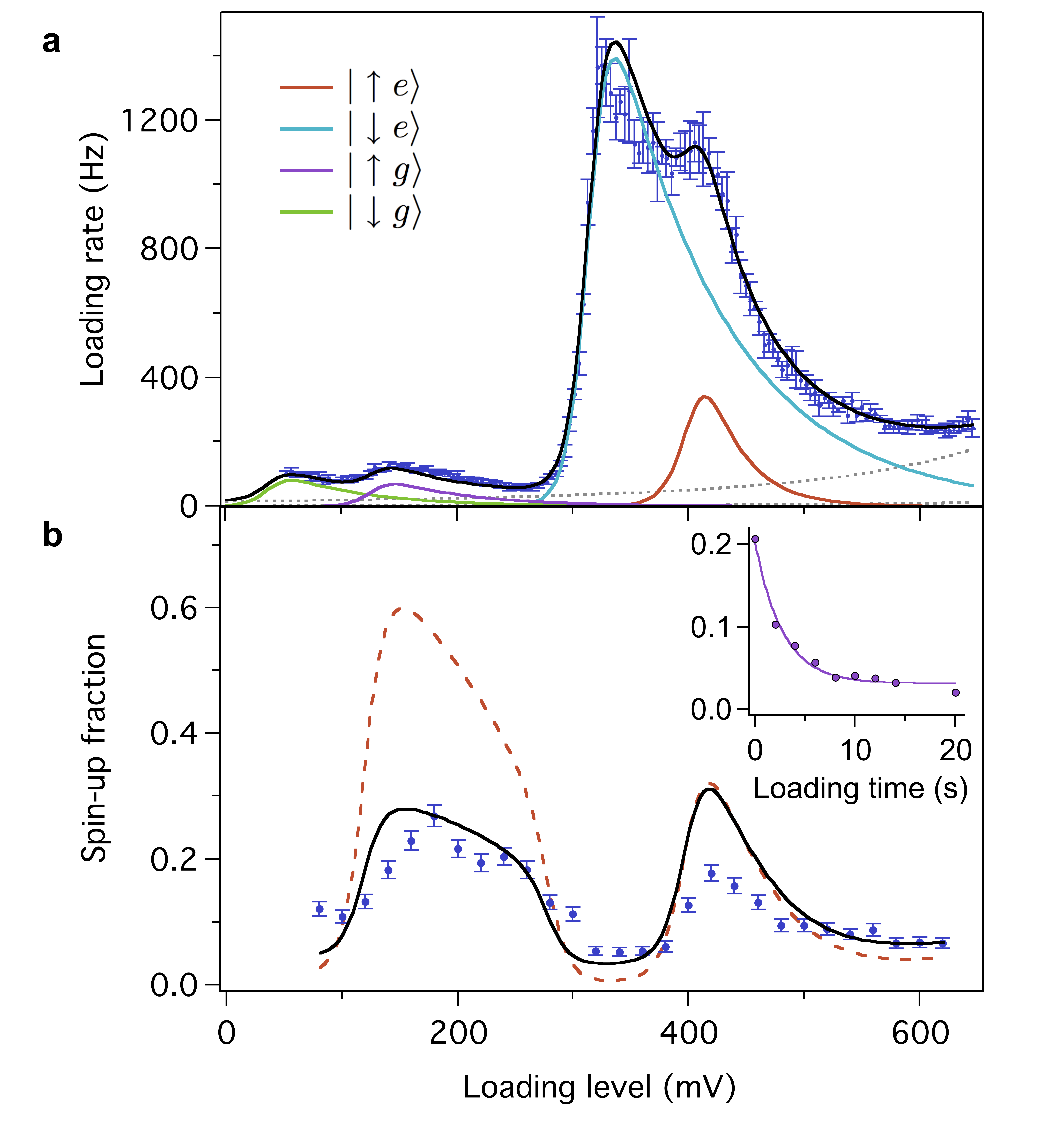}
\caption{\label{fig:data}
\textbf{Spin selective loading and spin lifetime measurement.} 
\textbf{a} Plot of electron loading rate \sym{\Gamma}{load} versus loading voltage \sym{V}{load} on gate PL at \amount{B=1.85}{T}.  Error bars are the standard deviation of four measurements.  Peaks occur when levels in the dot are made available for loading, enabling tuning of both the loaded-spin fraction and the loading rate.  The spin-split excited orbital levels \upe~and \downe~load \amount{\sim10}{} times faster than the ground state.  The black solid line in \textbf{a} is a fit obtained by treating the loading rate contribution from each state as a convolution of a Fermi-Dirac distribution,  Lorentzian lineshape, and a linearized energy dependent tunneling function.
\textbf{b} Spin-up fraction versus loading level measured using single-shot readout (see Fig.~\ref{fig:measurement}), with \amount{\sym{t}{load}=100}{ms}, \amount{\sym{t}{readout}=200}{ms}, \amount{\sym{t}{empty}=50}{ms}, \amount{\sym{V}{readout}=+85}{mV} and \amount{\sym{V}{empty}=-200}{mV}.  
Each data point corresponds to an analysis of 1000 single-shot traces.
The red dashed line is a fit using the same parameters as in \textbf{a};
the black solid line is the same fit with corrections for the $T_1$ decay before measurement, a measurement fidelity less than unity, and a non-zero dark count rate.  Error bars are $\sqrt{M}$, where $M$ is the number of readout events, and details of the fitting can be found in the Supplemental information.
\textbf{Inset:} Spin-up fraction versus loading time with \amount{\sym{V}{load}=+425}{mV}.  The solid line is an exponential fit that yields \amount{T_1= 2.8 \pm 0.3}{s}.  
}
\end{figure}

The rate at which an electron loads into the dot, as well as the spin state of that electron, both depend strongly on the voltage at which the electron is loaded.  In Fig.~\ref{fig:data}, we show that these features are observed over a wide range in gate voltage. Fig.~\ref{fig:data}a shows that the loading rate \sym{\Gamma}{load}, measured in the same way as in Fig.~\ref{fig:measurement}d--f, but now as a continuous function of loading level, increases by over an order of magnitude when the electron is loaded through an orbital excited state instead of through the orbital ground state.  The loading rate shows two strong peaks as a function of loading voltage, corresponding to the excited states \upe~and \downe.  Because orbital relaxation is very fast compared to the experimental
time scale~\cite{Khaetskii:2000p12369,Friesen:2004p63}, one expects any electron loaded into
a spin-\up state to be measured in the state \upg, and any electron loaded into a spin-\down
state is measured in the state \downg.

The relative fraction of spin-\up and spin-\down loaded can be tuned, because the loading rate for each spin type is voltage dependent.  We demonstrate this directly by using single-shot spin readout to measure the spin-\up fraction as a function of the loading level, as shown in Fig.~\ref{fig:data}b.  The fraction of spin-\up electrons has two clear peaks as a function of the loading voltage, one corresponding to \upg~and a second to \upe.  These peaks arise because loading rates into specific spin states in the dot are a maximum when the state is near resonance with the Fermi level.

The variation in both the total loading rate and the spin-\up fraction can be understood
by calculating the loading rate for each spin state.  The probability of loading a spin-\up electron is $\Gamma_{\uparrow}/\sym{\Gamma}{load}$, where $\Gamma_{\uparrow}$ is the total rate for all spin-\up channels, and $\sym{\Gamma}{load}$ is the sum of the rates for all ways to load the dot. A global fit to the data in terms of loading through spin-split ground and excited states with Lorentzian broadening is shown in Fig.~\ref{fig:data}a and b.  The fit is to both the total loading rate and the fraction of spin-\up electrons.  We treat the loading rate contribution from each state as a convolution of a Fermi-Dirac distribution, a Lorentzian lineshape, and a linearized energy dependent tunneling function~\cite{Datta:1995}:
\begin{equation}\label{eq:fit}
\Gamma_{i}(V) = 
\int_{-\infty}^\infty \left( \frac{\Gamma_{0i} e^{-\alpha y/E_{i} }}{1+e^{-\alpha y/k_{B}T}} \right)     \frac{\gamma_i/2\pi}{(\gamma_i/2)^2 + (\delta V_i -y)^2}  dy ,
\end{equation}
where $E_{i}$ is the linearized energy dependent tunneling coefficient that relates to the transparency of the barrier~\cite{MacLean:2007p1499}; 
$\gamma_{i}$ is the full width at half maximum of the Lorentzian;
$\Gamma_{0i}$ is the amplitude;
$V$ is the loading voltage;
and $\delta V_i = V  - V_{i}$, where $V_{i}$ is the position of the state (see Supplemental information for additional details).  The fit is in good agreement with the experimental measurements for the loading rates and the fractions of spin-\up and spin-\down electrons, supporting the interpretation of the data in terms of loading through specific spin states of the dot.

The black line in Fig.~\ref{fig:data}b is corrected for the finite spin lifetime $T_1$ of the electrons loaded into the dot.  We determine $T_1$ by measuring the spin-\up fraction as a function of the duration of the loading pulse, \sym{t}{load}~\cite{Elzerman:2004p431}.  The inset of Fig.~\ref{fig:data}b shows a typical result for a magnetic field of \amount{1.85}{T}, with an exponential fit yielding the value \amount{T_1 = 2.8 \pm 0.3}{s}.  For these data, \amount{\sym{V}{load}=+425}{mV}, so that loading occurs predominantly through the excited orbital states.   
We have measured $T_1$ at different loading voltages and also measured $T_1$ for a
four-step pulse sequence in which the dot is held at a fourth ``wait'' voltage.
The values of $T_1$ were all of order seconds, except when both the ``load" and ``wait"
voltages were set so the state \upg~ was aligned just below $E_F$, where the value of $T_1$
is reproducibly much shorter: \amount{T_1 = 136 \pm 22}{ms}. 
In principle, virtual hopping of electrons from the dot to the leads can limit $T_1$~\cite{Vorontsov:2008:p226805}, but the predicted magnitude is much too small to explain the observed effect, and this mechanism is not consistent with the behavior when the ``wait" voltage is varied.  The possibility that the excited orbital states are long-lived is ruled out by performing single-shot measurement with the readout level positioned such that \downg~and \upg~are below $E_F$ and \downe~and \upe~are above $E_F$.  In this situation, even for short loading times (\amount{\sym{t}{load} = 20}{ms}), less than 0.5\% of traces show events.

It has been observed that $T_1$ can be limited by dipolar coupling to nearby spins \cite{Morello:2010p2645}; we believe that the shorter $T_1$ time measured when the dot is loaded through the state \upg~is due to interaction with a nearby spin trap, which is suppressed by the deeper pulse associated with loading at the excited orbital states.  The ability to load an electron into the dot through the excited state thus provides an
immediate benefit, in that the additional tunability enables one to avoid sample imperfections that lead to a shorter spin relaxation time.

We have demonstrated the ability to manipulate and measure the spin state of individual electrons in a Si/SiGe quantum dot, and report the first single-shot measurements of the longitudinal spin relaxation time \sym{T}{1} in such devices.  The stability and tunability of the dot enable tunable loading rates and tunable spin loading.  These capabilities enhance the prospects for the development of Si/SiGe devices for spintronics and quantum information processing applications.

\section*{Methods Summary}
Fabrication method:  The dot was fabricated from a Si/SiGe heterostructure using the methods outlined in Ref.~\cite{Thalakulam:2010p183104}.  Magnetoresistance measurements using a Hall bar fabricated on the same heterostructure gave a carrier density of \amount{4\times10^{-11}}{cm^{-2}} and a mobility of \amount{50,000}{cm^2/Vs}.  The Pd top-gates were defined by electron beam lithography in the pattern shown in Fig.~\ref{fig:expt}a.  Measurements were performed in a dilution refrigerator with a base temperature of \amount{\sim20}{mK}.  The electron temperature was \amount{143 \pm 10}{mK}, measured by fitting the width of electronic transitions as the base temperature of the dilution refrigerator was raised.  The dot was tuned to the few-electron regime, and we estimate that we measured transitions between two-and three-electron occupation of the dot. Two of the electrons formed a ``closed shell"~\cite{Kouwenhoven:1997p1788} and were essentially inert, and our measurements were sensitive to the state of the third electron moving into and out of the dot.

Experimental method:  Pulsed gate voltages were supplied using a Tektronix AFG3022B arbitrary waveform generator.  The tunnel barriers of the dot were tuned so that the tunnel rates are in the range 10-1000 Hz, substantially slower than the inverse of the rise time of the pulse generator, which was $\leq 18$~ns.  The time resolution of the data acquisition was approximately 1 ms; it was limited by the bandwidth of the current preamp (DL Instruments Model 1211), which in turn was limited by the need to avoid ringing in the amplification circuit.  While the spin relaxation time can be detected using time-averaged measurements \cite{Xiao:2010p1876,Hayes:2009p1853}, single-shot measurement provides a direct measurement of the spin-\up fraction.  Identification of loading and unloading events during the single-shot readout phase was performed by wavelet analysis (see Supplemental information for details and for comparison to thresholding analysis).  The charge sensing signal from the low-frequency quantum point contact (QPC) corresponds to \amount{\sim 1 \times 10^{-3}}{e^2/h}.

\section*{Acknowledgments}
We acknowledge useful conversations with C. Tahan and M. Thalakulam, and experimental assistance from B. Rosemeyer and D. Greenheck.  This work was supported in part by ARO and LPS (W911NF-08-1-0482), by NSF (DMR-08325634, DMR-0805045), and by DARPA (N66001-09-1-2021).
The views, opinions, and findings contained in this article are those of the authors and should not be interpreted as representing the official views or policies, either expressed or implied, of the Defense Advanced Research Projects Agency or the Department of Defense.
This research utilized NSF-supported shared facilities at the University of Wisconsin-Madison.
The authors declare that they have no competing financial interests.
Correspondence and requests for materials should be addressed to M.A.E.~(maeriksson@wisc.edu).

\begin{onecolumngrid}

\newcounter{subequation}
\renewcommand{\theequation}{S\arabic{subequation}}
\setcounter{subequation}{1}

\newcounter{subfigure}
\renewcommand{\thefigure}{S\arabic{subfigure}}
\setcounter{subfigure}{1}

\newcounter{subtable}
\renewcommand{\thetable}{S\arabic{subtable}}
\setcounter{subtable}{1}

\section*{Supplemental Information}

\section{Analysis of real-time electron counting data.}

All of the single-shot counting experiments in the paper required detecting single electron tunneling events, as indicated by changes in the quantum point contact (QPC) current \sym{I}{QPC}.  Two different algorithms were employed for this determination.  The first algorithm used wavelets to determine transition edges and, subsequently, levels in the QPC signal.  All of the data in the paper were analyzed with this wavelet algorithm.  In the second algorithm, the data were inspected using a thresholding technique. Instead of detecting edges of transitions in the signal, this method compares the value of the QPC signal to a threshold value and thereby infers the state of the dot.  Both algorithms are described in greater detail below, and we show that the results of the two approaches to the analysis are consistent. 

\subsection{Electron tunnel event detection using wavelets}

A single electron tunneling onto or off of the quantum dot causes changes in the magnitude of \sym{I}{QPC}, leading to steps or ``edges'' in $\sym{I}{QPC}(t)$.  Wavelet analysis is a useful tool to determine the significance of edges.  In fact, the classic Canny edge detection algorithm \cite{Canny86} can be interpreted in the context of wavelets.  Based on observation of the real wavelet transform of our data and its noise characteristics, a robust edge detection algorithm was constructed based on the Stanford Wavelab library \cite{wavelab}.  The algorithm has strong similarities to the Canny algorithm.  First derivative of Gaussian wavelets \cite{waveletsbook} are used, implemented under the name `DerGauss' in Wavelab. Significant edges are indicated by (i) the presence of a modulus maxima track over many scales of the wavelet transform, and (ii) the amplitude of the wavelet transform ($A_W$) at the location of its track \cite{waveletsbook}.

The design of the algorithm and the tuning of its parameters were initially accomplished by the selection of a set of single-shot traces displaying a wide variety of inputs.  These inputs included traces with events on short, middle, and long time scales; traces with no, few, or several events; and traces with a variety of significant features that did not correspond to actual events.  The wavelet transforms of all traces in this set were observed, which allowed the construction of the algorithm and the initial tuning of detection parameters.  Subsequently, a few thousand traces were analyzed and verified by hand, and parameters were tweaked to approach a detection rate of 98\%.  It should be noted that, in all steps of the algorithm but the last, we try to err in the direction of selecting too many (possibly spurious) edges, waiting until the last step to reject spurious events.

The sampling rate employed in this experiment was \amount{40}{kHz}, much higher than the \amount{1}{ms} rise time of the current amplifier.  To speed up processing, bins of ten data points were averaged, which did not result in a loss of significant features.  The `DerGauss' real wavelet transform was computed using four scales per octave, which provides adequate resolution in the wavelet domain to make real features stand out visually.  The modulus maxima of this wavelet transformation is then computed.  A sample plot of a wavelet transformation with its modulus maxima superimposed can be seen in Fig.~\ref{fig:wavelet_thresh}c for the trace shown in a.

%Figure S1
\begin{figure}
\centering
\includegraphics[width=4.8in]{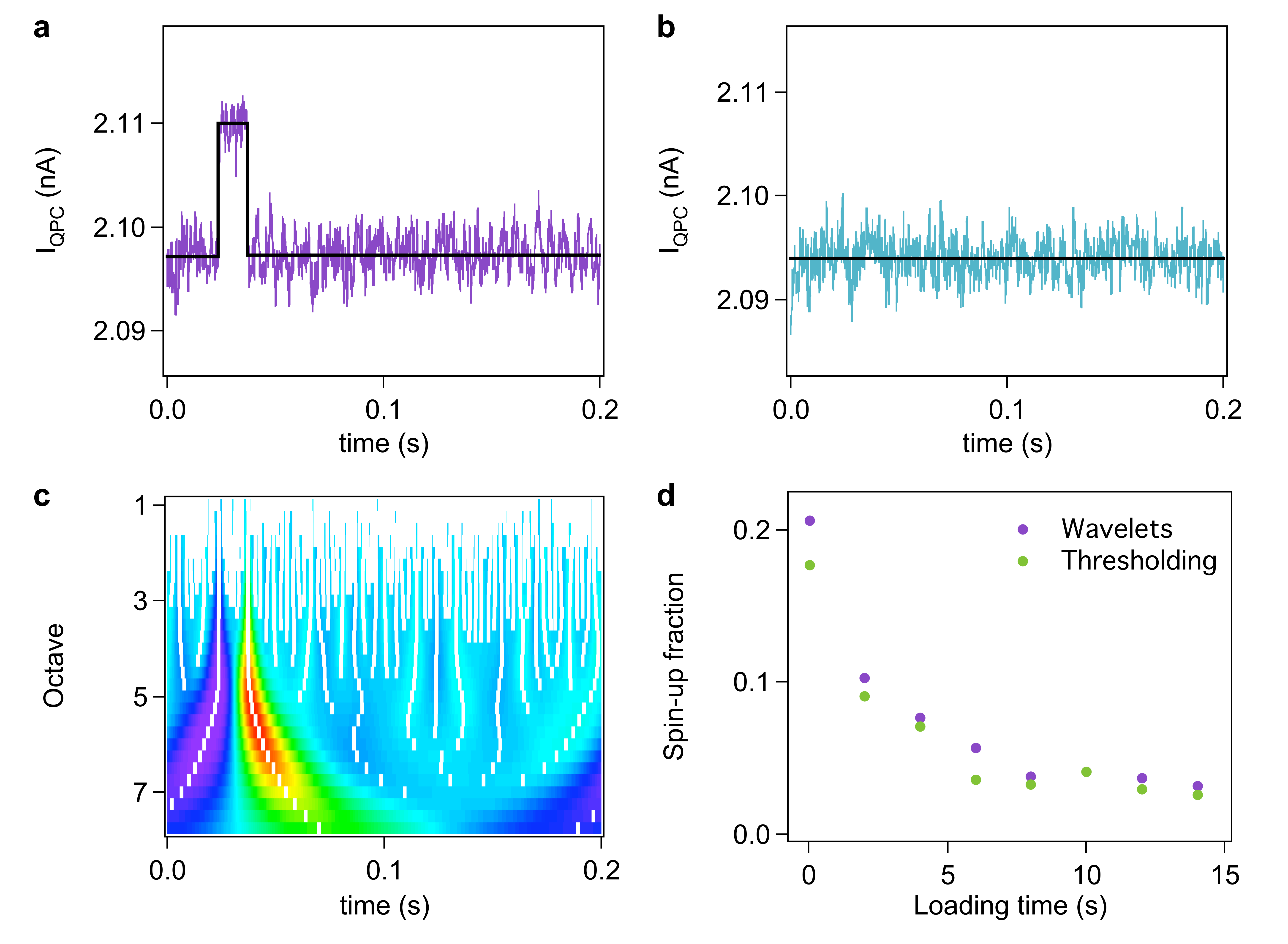}
\caption
{\textbf{Single-shot trace analysis using wavelets and thresholding.} \textbf{a \& b,} Raw data and reconstructed signals (shown in black) using the real wavelet transform for the two single-shot traces presented in Fig.~2a, b of the main text. \textbf{c,} Real wavelet transform and modulus maxima (shown in white) for the trace shown in \textbf{a}.  \textbf{d,} Results of the $T_1$ measurement presented in the inset to Fig.~3b of the main text, analysed using both wavelet and thresholding algorithms. Exponential fits yield $T_1=2.75 \pm 0.33$~s for the wavelet analysis results and $T_1=2.72 \pm 0.37$~s for the thresholding analysis results.}
\label{fig:wavelet_thresh}
\end{figure}
\addtocounter{subfigure}{1}  

The algorithm then iterates from the coarsest to the finest scale of the wavelet transform.  At each scale, the wavelet amplitude squared ($A_W^2$) for each modulus maxima track is computed.  Tracks for which this value reaches ten times the median value of all tracks at that scale are stored for further processing.  The purpose of this step is to select edges that are significantly larger than all other edges at the time scale under examination, which is reasonable because few events are expected in any particular trace.  For the results presented in this paper, increasing or decreasing the cutoff value by a factor of two results in less than a 5\% change in the correct detection ratio. 

The stored tracks are then followed to the next finer time scale, a procedure which is repeated until the best estimate of position for an edge is obtained at the finest time scale.  For each modulus maxima track this is accomplished by finding the closest track for which the wavelet amplitude has the same sign on both scales.  Some tracks representing spurious features, such as an insignificant change in the base level of the signal, or the remainder of a crosstalk edge at the beginning or end of the data, may fall off the edge of the wavelet transform.  For this reason, all tracks that come to within 2.5 rise times of the edge are rejected.  If cross talk from the applied pulse sequence were not significant, this parameter could be set much less aggressively, as in that case only tracks that truly fall off the edge need to be rejected.  Halving or doubling this parameter causes up to a 10\% change in correct detections.  At each time scale in the iteration, new tracks that suddenly reach the cutoff discussed in the previous paragraph are stored and considered in all subsequent iterations.  For each track followed, $A_W^2$ is summed from the scale at which it is detected until the finest scale.

When the finest time scale is reached, any edges for which the cumulative $A_W^2$ does not meet a secondary cutoff are rejected.  This cutoff is set by finding the typical maximum value of $A_W^2$ for a feature of interest by hand for a few traces, and requiring a sum of at least four times this amplitude.  This corresponds to a significant presence of the edge over at least one octave.  This will accept, for example, short tracks that are very strong, such as sharp spikes in the signal, but also longer tracks that are a bit weaker, such as long pulses with more gently sloping edges.  Halving or doubling this parameter causes approximately a 10\% change in the correct detection ratio  

All cutoffs up to this point were set to risk acceptance of edges of spurious features rather than to risk rejection of edges of weaker, true features.  The last step in the algorithm calculates the levels between edges found, and uses this information to perform a final rejection based on the physics of interest.  This step can have a large impact on which features are detected. While the algorithm used for this last step works well for the data in this paper, a completely different approach may be necessary for data with vastly different noise or signal characteristics.  For these particular experiments, edges that do not border significant changes in \sym{I}{QPC} are rejected.  This is accomplished by rejecting level changes that are less than five times the standard deviation of the signal between all edges.  This standard deviation is a good measure of high frequency noise present in the signal.  The sensitivity to this parameter is rather large, resulting in up to a 30\% change in correct detection ratio when the parameter is increased or decreased by one standard deviation, but once set, it is robust to the variation in high frequency noise encountered in the extensive quantity of data analyzed for the current experiments. 

Figure~\ref{fig:wavelet_thresh}a and b show results from using this wavelet edge detection algorithm for the readout phase of the two single-shot traces presented in Fig.~2a, b of the main text.  The reconstructed signals using the real wavelet transform are shown in black superimposed on the data.  An up-edge and a down-edge are detected in a, corresponding to the unloading of a spin-\up electron followed by the reloading of a spin-\down electron, and no edges are detected in b.

\subsection{Electron tunnel event detection using thresholding}

In this algorithm, the QPC signal is compared to a threshold. A tunneling event is detected if \sym{I}{QPC} crosses the threshold. If the signal remains below the threshold, zero tunneling events are assumed. While simple in concept, the thresholding analysis is complicated by the need to account for signal drift and high frequency noise.

First, the data are corrected for slow drifting of the QPC signal by subtracting the value of \sym{I}{QPC} corresponding to the electron located on the quantum dot (loaded) from every trace in a given experimental run. The value to subtract from each trace is determined in one of two ways, depending on the variance of the signal. If the variance is small (as for traces with zero tunnel events), then the loaded QPC signal is determined as the mean of the data. If the variance is large (as for traces containing tunneling events), then a histogram of the data is produced. This histogram contains a noise-broadened double peak, with the peaks centered on the values of \sym{I}{QPC} corresponding to the electron on or off the quantum dot. Least-squares fitting of a sum of two Guassians identifies the centers of the peaks, and the lower of the two center values is taken as the loaded \sym{I}{QPC} value.

After the drift correction, the threshold value can be determined.  A combined histogram of the QPC signal for many single-shot traces is produced. This histogram contains a double peak where, again, the two peaks are centered on the loaded and unloaded values of \sym{I}{QPC}. The histogram is fit to a sum of two Guassian peaks, and the threshold value is chosen as the location of the minimum between the fitted peaks. 

Once the threshold is established, every single-shot trace can be identified as containing tunnel events (the signal crosses the threshold), or containing no tunnel events (the signal remains below the threshold).  To reduce the probability that short, noise-induced spikes in the signal are misidentified as real tunnel events, only above-threshold signals that last longer than twice the time constant of measurement amplifier are counted.

\subsection{Comparison of wavelet and thresholding analysis}

All relevant data were analyzed employing both wavelet and thresholding methods.  The results show good quantitative agreement (see Figure~\ref{fig:wavelet_thresh}d). The advantage of the wavelet algorithm is that it allows each trace to be interpreted independently of all other traces. It is therefore uniquely suited for live, real-time analysis.

%%%%%

\section{Measurement of the gate voltage lever arms, the electron temperature and the Zeeman energy}

The spectroscopy in Fig.~1c,d of the main text show the electron response to an 800~kHz square wave pulse applied to gate L (see \cite{Elzerman:2004p731} for details of the spectroscopy technique).  The rate for an electron to tunnel onto the quantum dot is tuned to be slow compared to this pulse frequency.  A lock-in signal with a carrier frequency of 97~Hz applied to gate L is used to measure $d\sym{I}{QPC}/d\sym{V}{L}$ in response to the square wave pulse.  The dark lines indicate when additional energy levels become accessible resulting in a sudden increase in the electron tunneling rate.  In both c and d, the dark contrast of the excited orbital state indicates that it is more strongly coupled to the reservoir than the ground orbital state, because it causes a larger increase in the electron tunnel rate and thus the electron response.  Only the lower energy of the strongly coupled orbital excited states, \downe, is clearly visible in this spectroscopy, because once this state is accessible the tunnel rate is comparable to the pulse frequency and additional energy levels have no effect on the electron response.

In order to translate the spectroscopy data into energy levels on the dot, it is necessary to determine the lever arm, $\alpha$, for gate L.  The width of a charge transition peak in gate voltage is determined by both $\alpha$ and the electron temperature \cite{Houten:2005p1388}.  In order to determine these variables independently, we fit the width of a charge transition as a function of temperature by increasing the temperature of the dilution refrigerator.  The results for fridge temperatures ($\sym{T}{fridge}$) ranging from 24~mK (base temperature) to 335~mK are plotted in Fig.~\ref{fig:TevsTf}a.  The peak indicating the charge transition clearly decreases in amplitude and increases in width as the temperature is increased.  

We extract the width of each peak in Fig.~\ref{fig:TevsTf}a in terms of $T_e/\alpha$ by performing a global fit to the data. Each peak is fit with the equation,
\begin{equation}
dI_{\text{QPC}}/dV_{\text{L}}(V_L) = \frac{B}{4 k_B T_e}\mathrm{cosh}^{-2} \left( \frac{\alpha(V_0-V_L)}{2 k_B T_e} \right) +D
\end{equation}
\addtocounter{subequation}{1}
where $B$, $D$ and $V_0$ are fit parameters that are the same for all the peaks, and $T_e/\alpha$ is a fit parameter that is optimized for each peak independently.  The fit results for $T_e/\alpha$ for each of the peaks in Fig.~\ref{fig:TevsTf}a are shown in b.  As expected, at high temperatures $T_e/\alpha$ is linear in $\sym{T}{fridge}$, while at low temperatures $T_e/\alpha$ saturates.  We fit this cross-over using the functional form $T_e=\sqrt{T_0^2+\sym{T}{fridge}^2}$.  This fit is shown as the solid blue line in Fig.~\ref{fig:TevsTf}b, and the results for the two fit parameters, $\alpha$ and $T_0$ give $\alpha=0.125 \pm 0.006$~eV/V for gate L and $T_e=143 \pm 10$~mK at base temperature.  

With a measure of the electron temperature, it is possible to determine the lever arm for additional gates by performing a single fit.  We used gate PL to sweep through the same charge transition at base temperature, and the resulting fit gives $0.00144 \pm 0.00013$~eV / V for $\alpha$ multiplied by the attenuation (note: the voltage on gate PL is attenuated by a factor of $\sim30$).

%Figure S2
\begin{figure}
\includegraphics[width=4.7in]{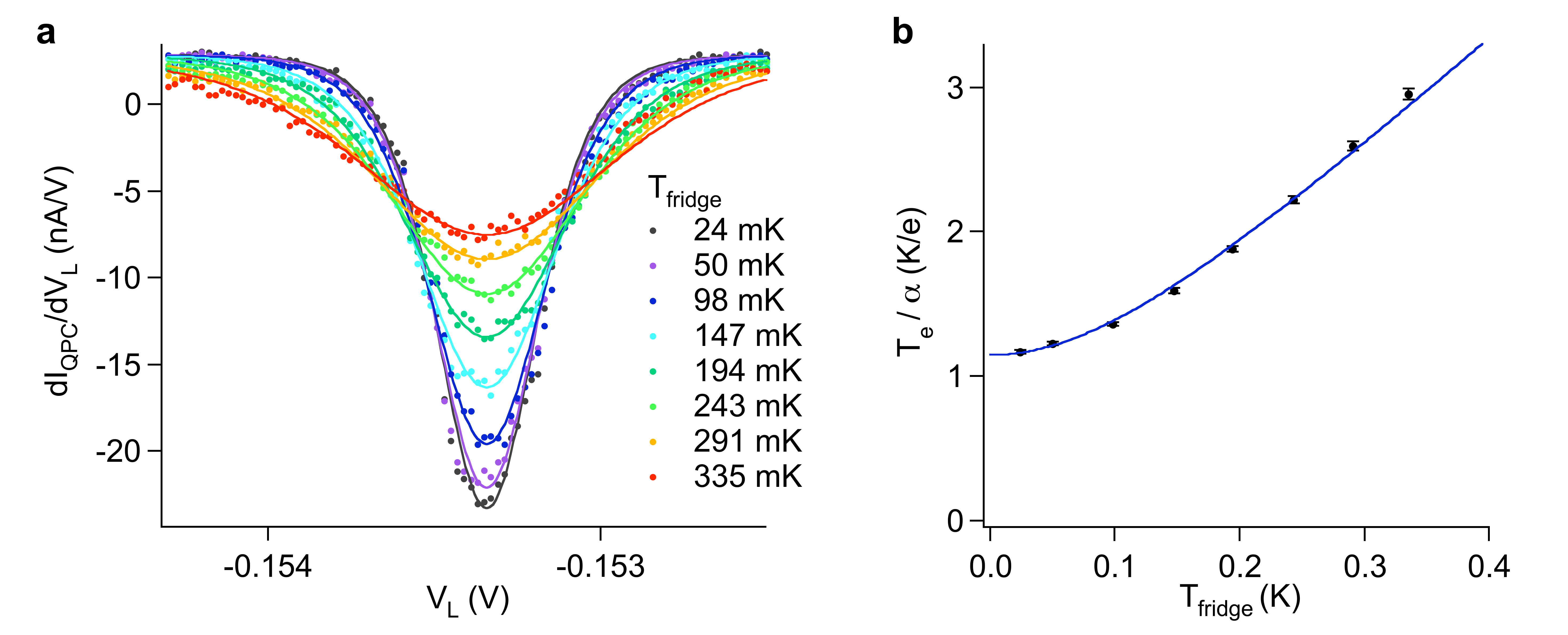}
\caption{\textbf{Measurement of $\alpha$ and electron temperature.} \textbf{a,} A charge transition peak measured with fridge temperatures ranging from 24~mK to 335~mK.  \textbf{b,} The width of the charge transitions for each of the temperature traces shown in \textbf{a} determined from a global fit. Error bars for each data point are determined from the uncertainties in the fit.}
\label{fig:TevsTf}
\end{figure}
\addtocounter{subfigure}{1}

Using the above results for $\alpha$ for gate L, it is possible to extract the Zeeman energy as a function of magnetic field, $B$, for the ground orbital state from the excited state spectroscopy data presented in Fig.~1d of the main text.  We take vertical line cuts to determine the separation, $\Delta V_L$, between the peaks corresponding to the states $| \downarrow g \rangle$ and $| \uparrow g \rangle$ for $B \approx 0.7$ to 2~T.  The results, after correcting for the gate voltage lever arm, are plotted in Fig.~\ref{fig:Zeeman}.

Similar to the results for a GaAs quantum dot presented in \cite{Hanson:2003p730}, at high magnetic fields we observe a clear deviation from the bulk silicon g-factor of 2 (black dashed line).  Following the method in \cite{Hanson:2003p730}, we fit the Zeeman splitting data with a second-order polynomial restricted to pass through the origin.  The result gives
\begin{equation}
g = (1.96 \pm 0.15) - (0.25 \pm 0.05)B(T).
\end{equation}
\addtocounter{subequation}{1}
The $B$-independent contribution of this result is in good agreement with the value for an electron spin in bulk silicon.

%Figure S3
\begin{figure}
\includegraphics[width=2.4in]{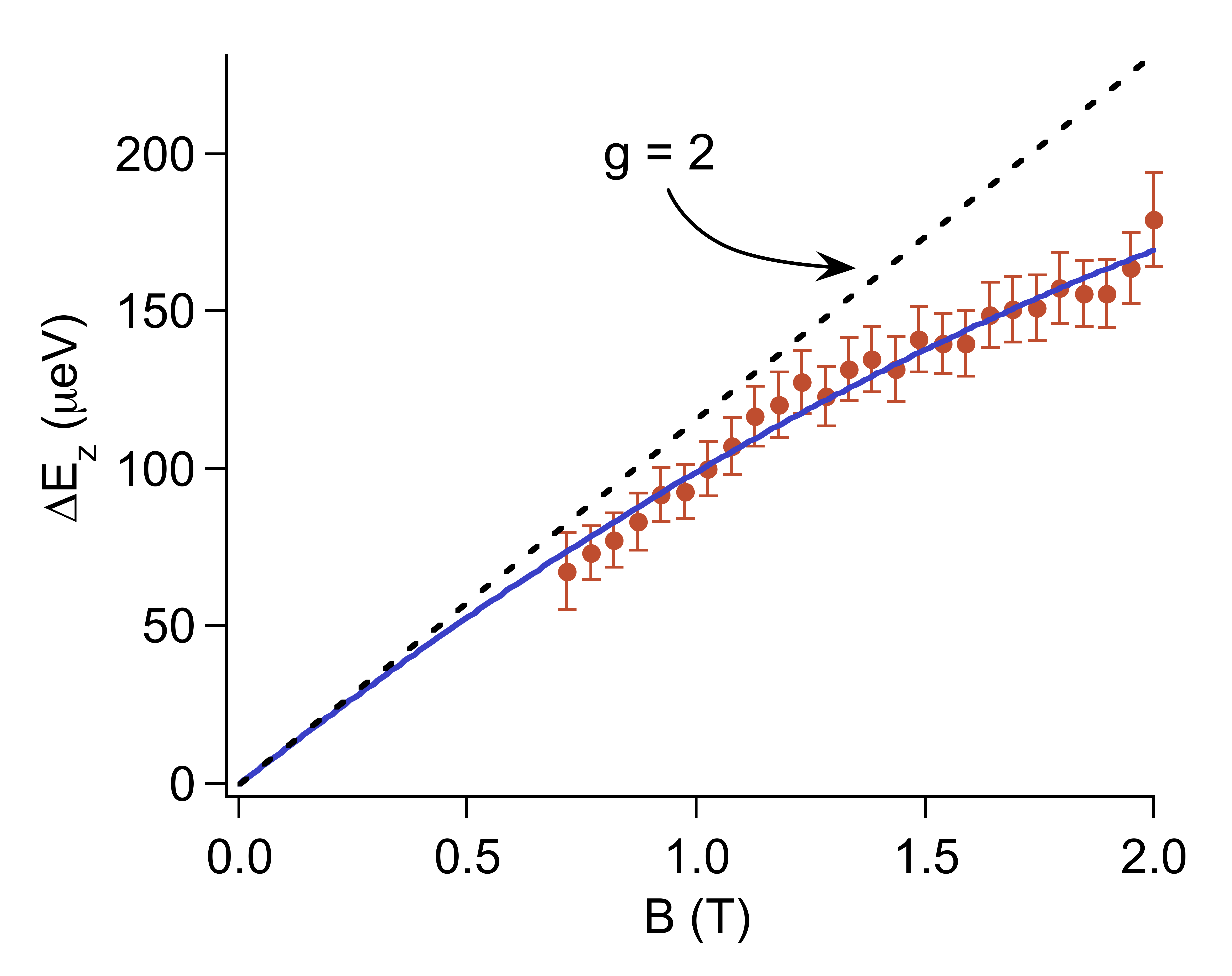}
\caption{\textbf{Zeeman splitting versus B-field.} Zeeman splitting as a function of magnetic field determined from the data presented in Fig.~1d of the main text.  Error bars are due to a combination of uncertainty in the peak separation, $\Delta V_L$, and uncertainty in the lever arm, $\alpha$ for gate L.  The dashed black line indicates $g=2$ and the solid blue line is the result of a fit using a second-order polynomial.}
\label{fig:Zeeman}
\end{figure}
\addtocounter{subfigure}{1}

%%%%%

\section{Global fit of loading rates and spin-\up fraction}

In this section, we give details of the modeling of the loading rates and the spin-\up fraction as a function of the loading gate voltage.  

\subsection{Modeling of the loading rates}

We perform a global fit to the data for loading rate and the spin-\up fraction as functions of the loading level, shown in Fig.~3a and b of the main paper.  Fig.~3a shows two pairs of peaks that correspond to the loading of the Zeeman-split ground \upg, \downg~and excited \upe, \downe~orbital states, as well as a shallow minimum on the right that we believe is due to a contribution from the loading of a higher orbital \uph, \downh. Fig. 3b shows two peaks as a function of loading level that correspond to the measurement of an increased fraction of spin-\up electrons. Tables \ref{tab:fit31} and \ref{tab:const} provide the numerical results and parameters for the global fit.

In fitting to the loading rate data of Fig. 3a, we take into account energy-dependent tunneling across the barrier~\cite{MacLean:2007p1499} as well as the lifetime broadening of energy states of the dot, and we use the Landauer-Buttiker formalism~\cite{Datta:1995} to model the loading rate. The lifetime broadening of energy levels is modeled using a Lorentzian function. The overall loading rate \sym{\Gamma}{load} as a function of loading level $V$ is given by $\sym{\Gamma}{load}(V)=\sum_{i}\sym{\Gamma}{i}(V)$, where the sum is over all possible transport channels $i \in \{\downarrow, \uparrow ; g,e, h\}$. 

For each of these channels, the loading rate is given by the convolution of the Fermi-Dirac distribution, the energy dependent tunneling function, and the Lorentzian function. It is given by

\begin{equation}\label{eq:fit}
\Gamma_{i}(V) = 
\int_{-\infty}^\infty \left( \frac{\Gamma_{0i} e^{-\alpha y/E_{i} }}{1+e^{-\alpha y/k_{B}T}} \right)     \frac{\gamma_i/2\pi}{(\gamma_i/2)^2 + (\delta V_i -y)^2}  dy ,
\end{equation}
where $E_{i}$ is the linearized energy dependent tunneling coefficient that relates to the transparency of the barrier~\cite{MacLean:2007p1499}; 
$\gamma_{i}$ is the full width at half maximum of the Lorentzian;
$\Gamma_{0i}$ is the amplitude;
$V$ is the loading voltage;
and $\delta V_i = V  - V_{i}$, where $V_{i}$ is the position of the state.
\addtocounter{subequation}{1}

In the fitting, the Zeeman splitting for each pair of states was identical, and initial guesses were made for the overall scale, position and Lorentzian width for the higher orbital states (\uph, \downh), with the leveling off of \sym{\Gamma}{load} at about \amount{600}{mV} as a guide.  These contributions are plotted as dotted lines in Fig. 3a of the main text. After each fitting run, the parameter estimates for \uph~and \downh~were revised such that the ratio of overall scaling of the rates and the fit repeated so that self-consistency is achieved.

Following the observations in Ref.~\cite{Amasha:2008p1500}, the amplitudes of the loading rates for different spin states were allowed to differ; we find, for each orbital, that $\Gamma_{\uparrow}<\Gamma_{\downarrow}$. 

We note that, while the energy dependent coefficients for the ground orbital states gave a good fit when they were constrained to be equal, this was not the case for the excited orbital states. It was necessary to have $E_{\downarrow e} > E_{\uparrow e}$, as can be seen from the fact that the downward slope on the right of the highest peaks, which is strongly dependent on $E_{\uparrow e}$, is much steeper than those of the ground orbital states.

The ground orbital states were modeled as delta functions, while the excited orbital states were modeled with a finite energy width as fit parameters, as consistent with the hypothesis of short-lived excited states. However, due to the fact that the temperature broadening of the Fermi function limits the resolution with which we can extract any energy information, only an upper bound on the energy width due to lifetime broadening could be obtained, giving a lifetime of at least $4\times10^2$ ps.

\subsection{Calculation of spin-\up fraction}

In the limit of infinite loading and unloading rates, and no spin-\up decay, the fraction, $n_\uparrow$, of spin-\up electrons measured as a function of loading level $V$ is simply
\begin{equation}\label{eq:simple}
n_\uparrow (V) = \frac{\Gamma_{\text{load}\uparrow} (V)}{\Gamma_{\text{load,total}}(V)},
\end{equation}
where the loading rates $\Gamma_i(V)$ are functions of the voltage $V$ at which the dot is loaded, $\Gamma_{\text{load}\uparrow}$ is the combined rate for loading all spin-\up channels, and $\Gamma_{\text{load,total}}$ is the total loading rate for all spin channels.  This quantity is plotted as the dashed red line in Fig.~3b of the main paper.

\addtocounter{subequation}{1}

So that the modeled spin-\up fraction correctly takes into account finite rates of loading, unloading and spin-\up decay, we set up rate equations for the load and readout phases and solve for the fraction of spin-\up at the end of each phase.

Given a pair of spin-\up and spin-\down states, the rate equations can be written as
\begin{eqnarray}
\frac{dn_{\uparrow}(t)}{dt} &=& -T_1^{-1} n_{\uparrow} + \Gamma_{\text{load}\uparrow} (1-n_{\uparrow}-n_{\downarrow}) - \Gamma_{\text{unload}\uparrow} n_{\uparrow} ,\\ 
	\addtocounter{subequation}{1}
\frac{dn_{\downarrow}(t)}{dt} &=& T_1^{-1} n_{\uparrow} + \Gamma_{\text{load} \uparrow} (1-n_{\uparrow}-n_{\downarrow}) - \Gamma_{\text{unload} \downarrow} n_{\downarrow} .
\end{eqnarray}
where $n_{\uparrow}(t)$ and $n_{\downarrow}(t)$ are the fractions of spin-\up and spin-\down respectively. They are explicit functions of time, and implicit functions of loading voltage $V$ via the loading rates $\Gamma_i(V)$.

\addtocounter{subequation}{1}
 
With the dot empty initially, we impose initial conditions $n_{\uparrow}(0) = n_{\downarrow}(0) = 0$ and solve to obtain the solution, which is a complicated function of the rates and time. However, during the load phase, both the spin-\up and spin-\down states are below the Fermi level of the lead, so the unloading rates are much smaller than the loading rates and can be neglected: i.e., $\Gamma_{\text{unload}\uparrow}=\Gamma_{\text{unload}\downarrow}=0$. This assumption leads to a simplified solution that offers more physical insight, given by
\begin{eqnarray}\label{eq:sol1}
n_{\uparrow}(t) &=& \frac{\Gamma_{\text{load}\uparrow}}{\Gamma_{\text{load}\uparrow}+\Gamma_{\text{load}\downarrow} -T_1^{-1} } e^{-t T_1^{-1}} (1-e^{-(\Gamma_{\text{load}\uparrow}+\Gamma_{\text{load}\downarrow} -T_1^{-1})t}) , \\
	\addtocounter{subequation}{1}
n_{\downarrow}(t) &=& \frac{ (\Gamma _{\text{load}\downarrow} - T_1^{-1})
   \left(1-e^{-t \left(\Gamma _{\text{load}\downarrow}+\Gamma _{\text{load}\uparrow}\right)}\right)+\Gamma _{\text{load}\uparrow} \left(1-e^{-t T_1^{-1}}\right)}{\Gamma _{\text{load}\uparrow}+\Gamma _{\text{load}\downarrow}-T_1^{-1}}.
\end{eqnarray}

\addtocounter{subequation}{1}

It is clear that, in the limit of loading rates fast compared to the decay rate, the approximation $n_\uparrow \approx e^{-t \Gamma{\text{decay}}} \left( \Gamma_{\text{load}\uparrow} / \Gamma_{\text{load,total}} \right)$ holds. If the rates are comparable, then the last term in brackets in Eq.~\ref{eq:sol1} contributes significantly.

We can generalize to include the loading of other spin states whose population comes from the tunneling in of electrons from the lead and not from the decay of the spin-\up state in question. This is necessary as the spin-\up states in question are each sandwiched between two spin-\down states and a higher orbital state. The solution to the spin-\up fraction then includes the total tunneling rate into all states, $\Gamma_{\text{total}}=\Gamma_{\text{load all}\downarrow} + \Gamma_{\text{load all}\uparrow}$, given by
\begin{equation}\label{eq:load}
n_{\uparrow} (t) = \frac{\Gamma_{\text{load}\uparrow}}{\Gamma_{\text{total}} -T_1^{-1} } e^{-t T_1^{-1}} (1-e^{-(\Gamma_{\text{total}} -T_1^{-1})t}).
\end{equation}

\addtocounter{subequation}{1}

Next, we consider the correction to the spin-\up fraction during the readout phase. During this phase, because the spin-\up state is above the Fermi level and the spin-\down state is below it, we neglect the spin-\up loading rate as well as the spin-\down unloading rate. Because the quantity measured in the experiment is the number of spin-\up electrons that tunnel out to the lead $n_{\text{c}\uparrow}$, we solve the following rate equations for $n_{\text{c}\uparrow}(t)$,
\begin{eqnarray}
\frac{dn_{\uparrow}(t)}{dt} &=& -T_1^{-1} n_{\uparrow} - \Gamma_{\text{unload}\uparrow} n_{\uparrow} ,\\
	\addtocounter{subequation}{1}
\frac{dn_{\downarrow}(t)}{dt} &=& T_1^{-1} n_{\uparrow} + \Gamma_{\text{load} \uparrow} (1-n_{\uparrow}-n_{\downarrow}), \\
	\addtocounter{subequation}{1}
\frac{dn_{\text{c}\uparrow} (t)}{dt} &=& \Gamma_{\text{unload}\uparrow}n_{\uparrow}.
\end{eqnarray}

The rate $\Gamma_{\text{unload}\uparrow}$ at which a spin-\up electron tunnels off the dot during the readout phase is determined by fitting to a histogram of the times at which up-edge is detected in the single-shot traces.  The results for the data presented in Fig. 3b of the main text are shown in Fig.~\ref{fig:unload}, where `0' corresponds to the start of the readout phase.  The exponential fit gives $\Gamma_{\text{unload}\uparrow}^{-1} = 21.6 \pm 1.3$ ms.

% Figure S4
\begin{figure}
\includegraphics[width=1.9in]{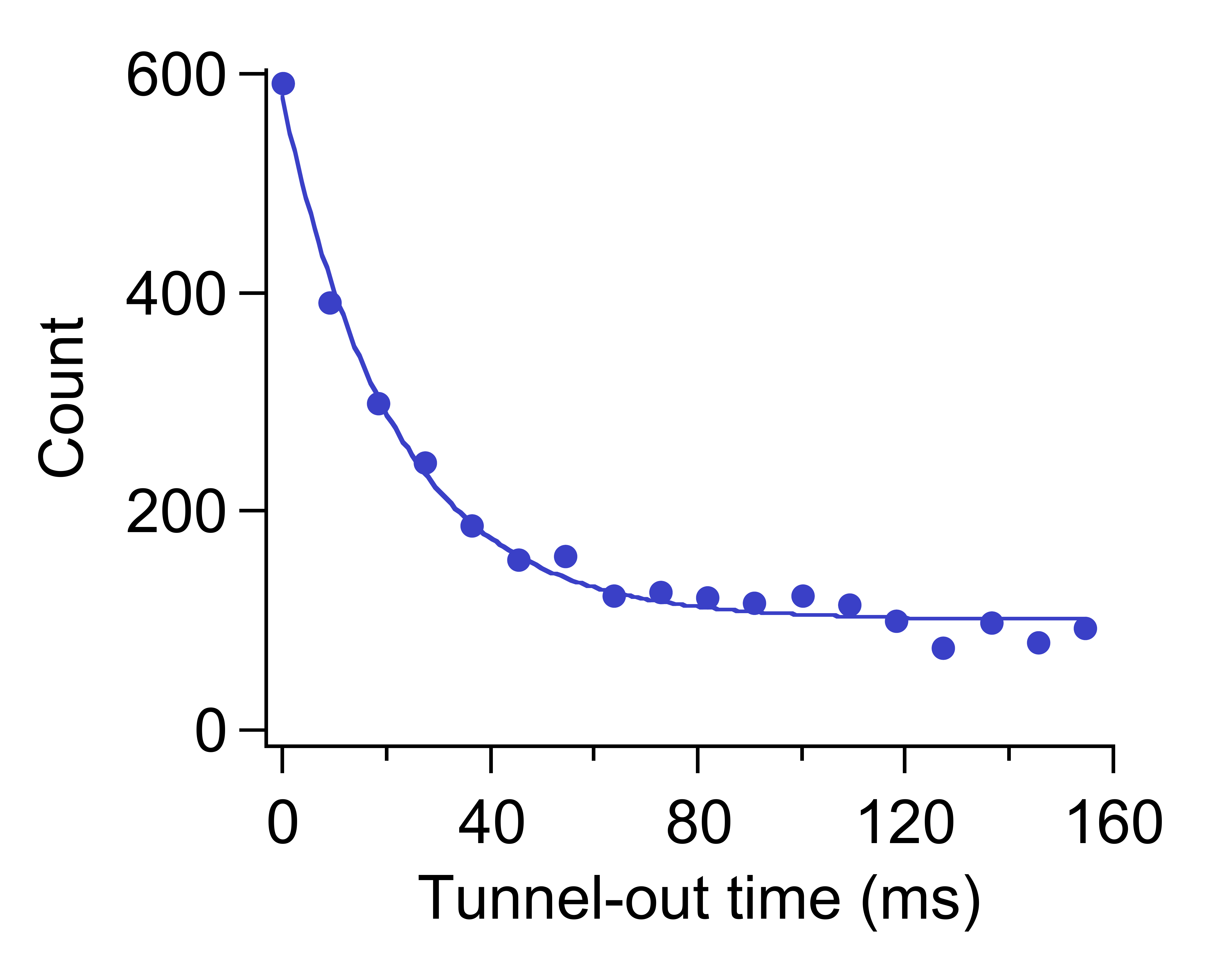}
\caption{\label{fig:unload}
\textbf{Measurement of the spin-\up unloading rate.} Histogram of the times at which an up-edge is detected in the readout phase of the single-shot traces for the data presented in Fig. 3b of the main text. The exponential fit gives the rate at which a spin-\up electron tunnels off the dot, $\Gamma_{\text{unload}\uparrow}^{-1}=21.6 \pm 1.3$ ms.
}
\end{figure}
\addtocounter{subfigure}{1}
\addtocounter{subequation}{1}

With the initial condition of a loaded spin-\up state, $n_{\uparrow}(0)=1$, we obtain the measured spin-\up fraction as
\begin{equation}
n_{\text{c}\uparrow} (t) = \frac{\Gamma _{\text{unload} \uparrow} }{T_1^{-1} + \Gamma _{\text{unload} \uparrow}} \left(1- e^{-t \left(T_1^{-1}+\Gamma_{\text{unload}\uparrow}\right)} \right)
\end{equation}

\addtocounter{subequation}{1}

With an independent determination of the spin-\up detection fidelity and the dark count rate ($92\%$ and $3\%$ respectively, as described in the next section), the measured spin-up fraction $\kappa_\uparrow$, fully corrected for finite tunnel rates during the load and readout phases of the 3-level pulse sequence and for spin-\up decay, is given by 
\begin{equation}
\kappa_\uparrow = 0.03 + 0.92\times n_{\uparrow}(t_{\text{load}}) \times n_{\text{c}\uparrow}(t_{\text{readout}}),
\end{equation}
where \sym{t}{load} and \sym{t}{readout} are the durations for the load and readout phases (100~ms and 200~ms respectively for the data in Fig.~3b). This result is plotted as the bold, black line in Fig.~3b of the main paper.

\addtocounter{subequation}{1}

\begin{table}
\caption{\label{tab:fit31}
Results of fit}
\begin{ruledtabular}
\begin{tabular}{ll}
Fit Parameter & Value  [Uncertainty]\\
\hline
Energy dependent coefficient \downg,~\upg~(meV) &	0.148 [27]  \\
Energy dependent coefficient \downe~(meV)	&	0.195 [30] \\
Energy dependent coefficient  \upe~(meV)	& 0.064 [23] \\
Position \upg~(meV)			&	0.186 [18] \\
Position \downe~(meV)			&	0.456 [42] \\
Position \upe~(meV)			&	0.585 [58] \\
Amplitude \downg~(Hz)		&	116 [8] \\
Amplitude \upg~(Hz)		&	110 [7] \\
Amplitude \downe~(Hz)		&	1,916 [62] \\
Amplitude \upg~(Hz)		&	610 [33] \\
Lorentzian width for \e~(meV)		&	5  $\times 10^{-6}$ \\

\hline
\underline{Quantities derived from fit parameters}&\\
Position \downg~(meV)			&	0.057 [9]  \\
Zeeman splitting (meV)					&	0.129 [17] \\
g-factor								&	1.20 [16] \\
Lifetime of \e~(ps)				&	440  \\
\end{tabular}
\end{ruledtabular}
\end{table}
\addtocounter{subtable}{1}  

% Table 2: Constants
\begin{table}
\caption{\label{tab:const}
Constants and independent parameters}
\begin{ruledtabular}
\begin{tabular}{lc}
temperature $T$ (mK)&143 [10]\\
magnetic field (T) & 1.85\\
alpha $\times$ attenuation (eV / V)& 0.00144 [13]\\
$T_{1,g}$ (ms) & 136 [22]\\
$T_{1,e}$ (ms) & $2.8 [3] \times 10^{3}$ \\
\hline
\underline{Parameters for \downh} &\\
Position (meV)&1.146\\
Lorentzian width  (meV)&0.433\\
Amplitude  (Hz) &3100\\
\hline
\underline {Parameters for \uph}&\\
Position (meV)&1.275\\
Lorentzian width (meV)&0.433\\
Amplitude  (Hz) &310\\
\end{tabular}
\end{ruledtabular}
\end{table}

\subsection{Single-shot spin readout visibility}

We determine the visibility of our spin readout using the method in \cite{Elzerman:2004p431}.  Since our g-factor is positive, spin-\down is the lower energy spin state in our case, and we therefore analogously define $\alpha$ as the probability that a spin-\down is detected as ``up" and $\beta$ as the probability that a spin-\up is detected as ``down", and we use corresponding definitions for the two contributions to $\beta$: $\beta_1$ and $\beta_2$.  The visibility of the spin readout is then given by $1-\alpha-\beta$.  

Due to the finite electron temperature, it is possible, although rare, for a spin-\down electron to tunnel off the dot during the readout phase even though its energy is below $E_F$, causing a spin-\down electron to be misidentified as spin-\up.  This `dark count' is determined by the saturation value of the exponential fit in the inset to Fig.~3b in the main text.  The result gives $\alpha = 0.031$.

During the readout phase, if a spin-\up electron tunnels off the dot and is replaced by a spin-\down electron at a rate comparable to the bandwidth of the measurement electronics, then the resulting step in \sym{I}{QPC} will be too small to detect, causing a spin-\up to be misidentified as spin-\down.  As in \cite{Elzerman:2004p431}, we determine the spin-\up detection fidelity by applying the 3-level pulse sequence in reverse, and we find it to be 91.8\%.  This gives $\beta_2=1-91.8\% = 0.082$. 

Due to the finite time for an spin-\up electron to tunnel off the quantum dot during the readout phase, it is possible that an electron that is spin-\up at the start of the readout might decay to a spin-\down before it tunnels off the dot.  This contribution to the spin readout fidelity is given by $\beta_1=1/(1+T_1\Gamma_{\text{unload}\uparrow})$.  From the exponential fit in Fig.~\ref{fig:unload}, we find $\Gamma_{\text{unload}\uparrow}^{-1}= 21.6$~ms.  This result combined with the measured value of $T_1 = 2.75$~s gives $\beta_1 = 0.0078$.   

Finally, we determine a spin-\down detection fidelity of $1-\alpha \approx 0.97$ and a spin-\up detection fidelity of $1-\beta \approx 0.91$, and this amounts to a total single-shot readout visibility of 88\%.  This visibility is notably improved from the 65\% reported in  \cite{Elzerman:2004p431}, and the majority of the improvement is due to the long spin lifetime compared to the relatively fast electron tunneling time.

\end{onecolumngrid}

\twocolumngrid

\bibliographystyle{naturemag}
\bibliography{silicon,preprints,wavelets}

\begin{thebibliography}{10}
\expandafter\ifx\csname url\endcsname\relax
  \def\url#1{\texttt{#1}}\fi
\expandafter\ifx\csname urlprefix\endcsname\relax\def\urlprefix{URL }\fi
\providecommand{\bibinfo}[2]{#2}
\providecommand{\eprint}[2][]{\url{#2}}

\bibitem{Appelbaum:2007p2644}
\bibinfo{author}{Appelbaum, I.}, \bibinfo{author}{Huang, B.} \&
  \bibinfo{author}{Monsma, D.}
\newblock \bibinfo{title}{Electronic measurement and control of spin transport
  in silicon}.
\newblock \emph{\bibinfo{journal}{Nature}} \textbf{\bibinfo{volume}{447}},
  \bibinfo{pages}{295--298} (\bibinfo{year}{2007}).

\bibitem{Kane:1998p133}
\bibinfo{author}{Kane, B.~E.}
\newblock \bibinfo{title}{A silicon-based nuclear spin quantum computer}.
\newblock \emph{\bibinfo{journal}{Nature}} \textbf{\bibinfo{volume}{393}},
  \bibinfo{pages}{133--137} (\bibinfo{year}{1998}).

\bibitem{Loss:1998p120}
\bibinfo{author}{Loss, D.} \& \bibinfo{author}{DiVincenzo, D.~P.}
\newblock \bibinfo{title}{Quantum computation with quantum dots}.
\newblock \emph{\bibinfo{journal}{Phys. Rev. A}} \textbf{\bibinfo{volume}{57}},
  \bibinfo{pages}{120--126} (\bibinfo{year}{1998}).

\bibitem{Vrijen:2000p1643}
\bibinfo{author}{Vrijen, R.} \emph{et~al.}
\newblock \bibinfo{title}{Electron-spin-resonance transistors for quantum
  computing in silicon-germanium heterostructures}.
\newblock \emph{\bibinfo{journal}{Phys. Rev. A}} \textbf{\bibinfo{volume}{62}},
  \bibinfo{pages}{012306} (\bibinfo{year}{2000}).

\bibitem{Friesen:2003p121301}
\bibinfo{author}{Friesen, M.} \emph{et~al.}
\newblock \bibinfo{title}{Practical design and simulation of silicon-based
  quantum-dot qubits}.
\newblock \emph{\bibinfo{journal}{Phys. Rev. B}} \textbf{\bibinfo{volume}{67}},
  \bibinfo{pages}{121301} (\bibinfo{year}{2003}).

\bibitem{Elzerman:2004p431}
\bibinfo{author}{Elzerman, J.~M.} \emph{et~al.}
\newblock \bibinfo{title}{Single-shot read-out of an individual electron spin
  in a quantum dot}.
\newblock \emph{\bibinfo{journal}{Nature}} \textbf{\bibinfo{volume}{430}},
  \bibinfo{pages}{431--435} (\bibinfo{year}{2004}).

\bibitem{Petta:2005p2180}
\bibinfo{author}{Petta, J.~R.} \emph{et~al.}
\newblock \bibinfo{title}{Coherent manipulation of coupled electron spins in
  semiconductor quantum dots}.
\newblock \emph{\bibinfo{journal}{Science}} \textbf{\bibinfo{volume}{309}},
  \bibinfo{pages}{2180--2184} (\bibinfo{year}{2005}).

\bibitem{Koppens:2006p766}
\bibinfo{author}{Koppens, F. H.~L.} \emph{et~al.}
\newblock \bibinfo{title}{Driven coherent oscillations of a single electron
  spin in a quantum dot}.
\newblock \emph{\bibinfo{journal}{Nature}} \textbf{\bibinfo{volume}{442}},
  \bibinfo{pages}{766--771} (\bibinfo{year}{2006}).

\bibitem{PioroLadriere:2008p776}
\bibinfo{author}{Pioro-Ladri\`{e}re, M.} \emph{et~al.}
\newblock \bibinfo{title}{Electrically driven single-electron spin resonance in
  a slanting {Z}eeman field}.
\newblock \emph{\bibinfo{journal}{Nat. Phys.}} \textbf{\bibinfo{volume}{4}},
  \bibinfo{pages}{776--779} (\bibinfo{year}{2008}).

\bibitem{Amasha:2008p1500}
\bibinfo{author}{Amasha, S.} \emph{et~al.}
\newblock \bibinfo{title}{Spin-dependent tunneling of single electrons into an
  empty quantum dot}.
\newblock \emph{\bibinfo{journal}{Phys. Rev. B}} \textbf{\bibinfo{volume}{78}},
  \bibinfo{pages}{041306} (\bibinfo{year}{2008}).

\bibitem{Elzerman:2004p731}
\bibinfo{author}{Elzerman, J.~M.}, \bibinfo{author}{Hanson, R.},
  \bibinfo{author}{van Beveren, L. H.~W.}, \bibinfo{author}{Vandersypen, L.
  M.~K.} \& \bibinfo{author}{Kouwenhoven, L.~P.}
\newblock \bibinfo{title}{Excited-state spectroscopy on a nearly closed quantum
  dot via charge detection}.
\newblock \emph{\bibinfo{journal}{Appl. Phys. Lett.}}
  \textbf{\bibinfo{volume}{84}}, \bibinfo{pages}{4617--4619}
  (\bibinfo{year}{2004}).

\bibitem{Tahan:2002p035314}
\bibinfo{author}{Tahan, C.}, \bibinfo{author}{Friesen, M.} \&
  \bibinfo{author}{Joynt, R.}
\newblock \bibinfo{title}{Decoherence of electron spin qubits in {S}i-based
  quantum computers}.
\newblock \emph{\bibinfo{journal}{Phys. Rev. B}} \textbf{\bibinfo{volume}{66}},
  \bibinfo{pages}{035314} (\bibinfo{year}{2002}).

\bibitem{Hanson:2007p1217}
\bibinfo{author}{Hanson, R.}, \bibinfo{author}{Kouwenhoven, L.~P.},
  \bibinfo{author}{Petta, J.~R.}, \bibinfo{author}{Tarucha, S.} \&
  \bibinfo{author}{Vandersypen, L. M.~K.}
\newblock \bibinfo{title}{Spins in few-electron quantum dots}.
\newblock \emph{\bibinfo{journal}{Rev. Mod. Phys.}}
  \textbf{\bibinfo{volume}{79}}, \bibinfo{pages}{1217--1265}
  (\bibinfo{year}{2007}).

\bibitem{Boykin:2004p115}
\bibinfo{author}{Boykin, T.~B.} \emph{et~al.}
\newblock \bibinfo{title}{Valley splitting in strained silicon quantum wells}.
\newblock \emph{\bibinfo{journal}{Appl. Phys. Lett.}}
  \textbf{\bibinfo{volume}{84}}, \bibinfo{pages}{115--117}
  (\bibinfo{year}{2004}).

\bibitem{Goswami:2007p41}
\bibinfo{author}{Goswami, S.} \emph{et~al.}
\newblock \bibinfo{title}{Controllable valley splitting in silicon quantum
  devices}.
\newblock \emph{\bibinfo{journal}{Nat. Phys.}} \textbf{\bibinfo{volume}{3}},
  \bibinfo{pages}{41--45} (\bibinfo{year}{2007}).

\bibitem{Kharche:2007p092109}
\bibinfo{author}{Kharche, N.}, \bibinfo{author}{Prada, M.},
  \bibinfo{author}{Boykin, T.~B.} \& \bibinfo{author}{Klimeck, G.}
\newblock \bibinfo{title}{Valley splitting in strained silicon quantum wells
  modeled with 2$\,^{\circ}$ miscuts, step disorder, and alloy disorder}.
\newblock \emph{\bibinfo{journal}{Appl. Phys. Lett.}}
  \textbf{\bibinfo{volume}{90}}, \bibinfo{pages}{092109}
  (\bibinfo{year}{2007}).

\bibitem{Lansbergen:2008p1545}
\bibinfo{author}{Lansbergen, G.~P.} \emph{et~al.}
\newblock \bibinfo{title}{Gate-induced quantum-confinement transition of a
  single dopant atom in a silicon {F}in{F}{E}{T}}.
\newblock \emph{\bibinfo{journal}{Nat. Phys.}} \textbf{\bibinfo{volume}{4}},
  \bibinfo{pages}{656--661} (\bibinfo{year}{2008}).

\bibitem{Culcer:2009p205302}
\bibinfo{author}{Culcer, D.}, \bibinfo{author}{Cywinski, L.},
  \bibinfo{author}{Li, Q.~Z.}, \bibinfo{author}{Hu, X.} \&
  \bibinfo{author}{{Das Sarma}, S.}
\newblock \bibinfo{title}{Realizing singlet-triplet qubits in multivalley {S}i
  quantum dots}.
\newblock \emph{\bibinfo{journal}{Phys. Rev. B.}}
  \textbf{\bibinfo{volume}{80}}, \bibinfo{pages}{205302}
  (\bibinfo{year}{2009}).

\bibitem{Friesen:2010p115324}
\bibinfo{author}{Friesen, M.} \& \bibinfo{author}{Coppersmith, S.~N.}
\newblock \bibinfo{title}{Theory of valley-orbit coupling in a {S}i/{S}i{G}e
  quantum dot}.
\newblock \emph{\bibinfo{journal}{Phys. Rev. B}} \textbf{\bibinfo{volume}{81}},
  \bibinfo{pages}{115324} (\bibinfo{year}{2010}).

\bibitem{Fuechsle:2010p502}
\bibinfo{author}{Fuechsle, M.} \emph{et~al.}
\newblock \bibinfo{title}{Spectroscopy of few-electron single-crystal silicon
  quantum dots}.
\newblock \emph{\bibinfo{journal}{Nat. Nanotechnol.}}
  \textbf{\bibinfo{volume}{5}}, \bibinfo{pages}{502--505}
  (\bibinfo{year}{2010}).

\bibitem{Khaetskii:2000p12369}
\bibinfo{author}{Khaetskii, A.} \& \bibinfo{author}{Nazarov, Y.}
\newblock \bibinfo{title}{Spin relaxation in semiconductor quantum dots}.
\newblock \emph{\bibinfo{journal}{Phys. Rev. B}} \textbf{\bibinfo{volume}{61}},
  \bibinfo{pages}{12639} (\bibinfo{year}{2000}).

\bibitem{Friesen:2004p63}
\bibinfo{author}{Friesen, M.}, \bibinfo{author}{Tahan, C.},
  \bibinfo{author}{Joynt, R.} \& \bibinfo{author}{Eriksson, M.~A.}
\newblock \bibinfo{title}{Spin readout and initialization in a semiconductor
  quantum dot}.
\newblock \emph{\bibinfo{journal}{Phys. Rev. Lett.}}
  \textbf{\bibinfo{volume}{92}}, \bibinfo{pages}{037901}
  (\bibinfo{year}{2004}).

\bibitem{Datta:1995}
\bibinfo{author}{Datta, S.}
\newblock \emph{\bibinfo{title}{Electronic Transport in Mesoscopic Systems}}
  (\bibinfo{publisher}{Cambridge University Press},
  \bibinfo{address}{Cambridge, UK}, \bibinfo{year}{1995}).

\bibitem{MacLean:2007p1499}
\bibinfo{author}{MacLean, K.} \emph{et~al.}
\newblock \bibinfo{title}{Energy-dependent tunneling in a quantum dot}.
\newblock \emph{\bibinfo{journal}{Phys. Rev. Lett.}}
  \textbf{\bibinfo{volume}{98}}, \bibinfo{pages}{036802}
  (\bibinfo{year}{2007}).

\bibitem{Vorontsov:2008:p226805}
\bibinfo{author}{Vorontsov, A.~B.} \& \bibinfo{author}{Vavilov, M.~G.}
\newblock \bibinfo{title}{Spin relaxation in quantum dots due to electron
  exchange with leads}.
\newblock \emph{\bibinfo{journal}{Phys. Rev. Lett.}}
  \textbf{\bibinfo{volume}{101}}, \bibinfo{pages}{226805}
  (\bibinfo{year}{2008}).

\bibitem{Morello:2010p2645}
\bibinfo{author}{Morello, A.} \emph{et~al.}
\newblock \bibinfo{title}{Single-shot readout of an electron spin in silicon}.
\newblock \emph{\bibinfo{journal}{Nature}} \textbf{\bibinfo{volume}{467}},
  \bibinfo{pages}{687--691} (\bibinfo{year}{2010}).

\bibitem{Thalakulam:2010p183104}
\bibinfo{author}{Thalakulam, M.} \emph{et~al.}
\newblock \bibinfo{title}{Fast tunnel rates in {S}i/{S}i{G}e one-electron
  single and double quantum dots}.
\newblock \emph{\bibinfo{journal}{Appl. Phys. Lett.}}
  \textbf{\bibinfo{volume}{96}}, \bibinfo{pages}{183104}
  (\bibinfo{year}{2010}).

\bibitem{Kouwenhoven:1997p1788}
\bibinfo{author}{Kouwenhoven, L.~P.} \emph{et~al.}
\newblock \bibinfo{title}{Excitation spectra of circular, few-electron quantum
  dots}.
\newblock \emph{\bibinfo{journal}{Science}} \textbf{\bibinfo{volume}{278}},
  \bibinfo{pages}{1788--1792} (\bibinfo{year}{1997}).

\bibitem{Xiao:2010p1876}
\bibinfo{author}{Xiao, M.}, \bibinfo{author}{House, M.~G.} \&
  \bibinfo{author}{Jiang, H.~W.}
\newblock \bibinfo{title}{Measurement of the spin relaxation time of single
  electrons in a silicon metal-oxide-semiconductor-based quantum dot}.
\newblock \emph{\bibinfo{journal}{Phys. Rev. Lett.}}
  \textbf{\bibinfo{volume}{104}}, \bibinfo{pages}{096801}
  (\bibinfo{year}{2010}).

\bibitem{Hayes:2009p1853}
\bibinfo{author}{Hayes, R.~R.} \emph{et~al.}
\newblock \bibinfo{title}{Lifetime measurements ({T}1) of electron spins in
  {S}i/{S}i{G}e quantum dots} (\bibinfo{year}{2009}).
\newblock \bibinfo{note}{ArXiv: 0908.0173v1}.

\bibitem{Canny86}
\bibinfo{author}{Canny, J.}
\newblock \bibinfo{title}{A computational approach to edge detection}.
\newblock \emph{\bibinfo{journal}{IEEE Transactions on Pattern Analysis and
  Machine Intelligence}} \textbf{\bibinfo{volume}{8}},
  \bibinfo{pages}{679--698} (\bibinfo{year}{1986}).

\bibitem{wavelab}
\bibinfo{author}{Buckheit, J.~B.} \& \bibinfo{author}{Donoho, D.~L.}
\newblock \bibinfo{title}{Wavelab and reproducible research}.
\newblock \bibinfo{pages}{55--81} (\bibinfo{publisher}{Springer-Verlag},
  \bibinfo{year}{1995}).

\bibitem{waveletsbook}
\bibinfo{author}{Mallat, S.}
\newblock \emph{\bibinfo{title}{A Wavelet Tour of Signal Processing}}
  (\bibinfo{publisher}{Academic Press}, \bibinfo{year}{1999}),
  \bibinfo{edition}{second} edn.

\bibitem{Houten:2005p1388}
\bibinfo{author}{van Houten, H.}, \bibinfo{author}{Beenakker, C. W.~J.} \&
  \bibinfo{author}{Staring, A. A.~M.}
\newblock \bibinfo{title}{Coulomb-blockade oscillations in semiconductor
  nanostructures}.
\newblock \emph{\bibinfo{journal}{arXiv}}
  \textbf{\bibinfo{volume}{cond-mat/0508454v1}} (\bibinfo{year}{2005}).

\bibitem{Hanson:2003p730}
\bibinfo{author}{Hanson, R.} \emph{et~al.}
\newblock \bibinfo{title}{Zeeman energy and spin relaxation in a one-electron
  quantum dot}.
\newblock \emph{\bibinfo{journal}{Phys. Rev. Lett.}}
  \textbf{\bibinfo{volume}{91}}, \bibinfo{pages}{196802}
  (\bibinfo{year}{2003}).

\end{thebibliography}

\end{document}